\RequirePackage{fix-cm}
\documentclass[smallextended,natbib]{svjour3}       
\smartqed  
\usepackage{graphicx}
\usepackage[colorinlistoftodos]{todonotes}

\usepackage{relsize}
\usepackage{subcaption} 
\usepackage{multirow}

\usepackage{lineno,hyperref}
\modulolinenumbers[5]

\usepackage{color}

\usepackage{amssymb}

\usepackage{amsmath} 

\usepackage[inline]{enumitem} 
\newcommand{\inlinelist}[2]{\begin{enumerate*}[label=#1] #2 \end{enumerate*}}

\def \figurename {Fig.}
\def \figurenamelong {Figure}
\newcommand{\figref}[1]{\figurename~\ref{#1}}
\newcommand{\Figref}[1]{\figurenamelong~\ref{#1}}

\def \tablename {Table}
\newcommand{\tabref}[1]{\tablename~\ref{#1}}

\def \sectionname {Section}
\newcommand{\secref}[1]{\sectionname~\ref{#1}}

\def \eqname {eq.}
\newcommand{\myeqref}[1]{\eqname~\eqref{#1}}

\def \eg {e.g., }
\def \ie {i.e., }

\def\RL{\textsf}

\def\e{\begin{equation}}
\def\f{\end{equation}}
\def\ea{\begin{eqnarray}}
\def\fa{\end{eqnarray}}

\newcommand{\run}[1]{\texttt{#1}}

\usepackage{mathabx}
\def\defis{\stackrel{\smalltriangleup}{=}}

\newcommand{\pRL}[2]{\ensuremath{p_{\RL{\scriptsize{#1}}\vert\RL{\scriptsize{#2}}}}}

\journalname{Information Retrieval Journal}
\begin{document}

\title{Predicting Relevance based on Assessor Disagreement: Analysis and Practical Applications for Search Evaluation%
}

\titlerunning{Predicting Relevance based on Assessor Disagreement}        

\author{Thomas Demeester  \and Robin Aly \\
        Djoerd Hiemstra \and Dong Nguyen \\
        Chris Develder
}


\institute{T. Demeester, C. Develder \at
              Ghent University - iMinds \\
              Belgium\\
              \email{\{tdmeeste, cdvelder\}@intec.ugent.be}
           \and
           R. Aly, D. Hiemstra, D. Nguyen \at
              University of Twente\\
              The Netherlands\\
              \email{\{r.aly, d.hiemstra, d.nguyen\}@utwente.nl}
}

\date{Received: 1 May 2015 / Accepted: 15 October 2015\\\\
The final publication is available at Springer via\\ http://dx.doi.org/10.1007/s10791-015-9275-x}

\maketitle

\begin{abstract}

Evaluation of search engines relies on assessments of search results for selected test queries, from which we would ideally like to draw conclusions in terms of relevance of the results for general (\eg future, unknown) users.
In practice however, most evaluation scenarios only allow us
to conclusively determine 
the relevance towards the particular assessor that provided the judgments. A factor that cannot be ignored when extending conclusions made from assessors towards users, is the possible disagreement on relevance, assuming that a single gold truth label does not exist. 
This paper presents and analyzes the Predicted Relevance Model (PRM), which allows predicting a particular result's relevance for a random user, based on an observed assessment and knowledge on the average disagreement between assessors. 
With the PRM, existing evaluation metrics designed to measure binary assessor relevance, can be transformed into more robust and effectively graded measures that evaluate relevance towards a random user. It also leads to a principled way of quantifying multiple graded or categorical relevance levels for use as gains in established graded relevance measures, such as normalized discounted cumulative gain (nDCG), which nowadays often use heuristic and data-independent gain values. Given a set of test topics with graded relevance judgments, the PRM allows evaluating systems on different scenarios, such as their capability of retrieving top results, or how well they are able to filter out non-relevant ones. 
Its use in actual evaluation scenarios is illustrated on several information retrieval test collections.

\end{abstract}

\section{Introduction}

Measuring the effectiveness of search results for users is essential for the improvement, comparison and tuning of search engines. To achieve this task, effectiveness measures often employ relevance labels assigned by assessors as ground truth. Hence, the literature often 
treats assessors as if they are actual users that pose a query on their current information need, and assess the returned results accordingly.
However, in practice assessors are often workers with the task to assess the relevance of results for users they have never met. To estimate the impact of this assumption, previous work studies the disagreement of assessors on binary labels and its influence on search engine comparisons \citep{Voorhees2001}, leading to the conclusion that search engine comparisons are stable even under substantial assessor disagreement. \citet{Demeester2014} show that in a graded relevance setting, this disagreement is especially strong on the top relevance levels. The current paper explicitly models the disagreement between assessors and particular scenarios of user relevance, such as users that are only satisfied with top results, or users that are looking for any result that is at least marginally relevant. This model is applied to graded relevance based evaluation. The findings are supported by experimental results on two different datasets.

Modeling the plurality of users involved in search currently receives much research interest. Most work focuses on differences among users, \eg diversity of search results~\citep{Zhai2003} and query ambiguity~\citep{Agrawal2009}. We propose that the plurality between users and assessors is equally important. So far, research on this topic considers deviations between assessors and users as mistakes, \eg due to input error or ambiguous instructions, and evaluation measures had to prove to be stable against these unwanted effects. In this paper we consider differences between assessors' predictions and users as natural and we propose methods to integrate them into effectiveness measures. 

We focus on exploiting our model for user-assessor plurality in standard evaluation measures based on graded assessment levels. Other measures, \eg for ambiguity, could also benefit from our model. However, their particular consideration of differences among users deviates from the presented approach, which makes the connection between assessor observations and user preferences difficult to isolate. 
In a standard graded relevance evaluation setup, the gains for the relevance grades often lack a direct connection to the assumed evaluation scenario and are typically set heuristically. For example, \citet{Kanoulas2009} set gain values of the normalized discounted cumulative gain (nDCG) effectiveness measure by optimizing formal quality criteria for test topics, but they do not model the connection with relevance towards users. The graded average precision (GAP) by \citet{Robertson2010} is one of the first to define gain values based on users: they consider user populations that perceive documents as relevant from specific thresholds in the ground truth relevance levels onwards. However, setting the GAP gains requires the hard task of determining the distribution of threshold values over the population. Our model also assumes a binary notion of relevance for individual users. However, it avoids the common assumption that there is a single ground truth, given by the labels assigned by the assessors. 

The main contribution of this paper is the Predicted Relevance Model (PRM), which captures differences between assessor judgments in order to estimate the relevance of documents for random users, with:
\begin{itemize}
 \item an assessor model with multiple relevance levels, 
 \item a user model based on binary relevance and linked to the assessor model, and
 \item a detailed estimation procedure of probabilities that quantify disagreement.
\end{itemize}

The PRM predicts the relevance for a random user, based on an observed assessment, and the expected relevance over different assessors. The model is built on the insights gained by~\citet{Demeester2014}, who introduced the User Disagreement Model (UDM). The main differences with respect to this previous work are:
\inlinelist{(\roman*)}{
\item refinements of the model, generalized with respect to only considering top relevance,
\item deeper analyses and insights into the model, 
\item new insights into applying the model on binary and graded relevance measures,
\item an experimental analysis of using the model for evaluating retrieval systems, and
\item a validation based on two different IR evaluation collections.%
}
The differences between the PRM and the UDM are discussed in more detail in \secref{subsec:UDM}.

Note that~\citet{Demeester2014} present evidence and a quantitative analysis of user disagreement, most of which will not be repeated here. For example, it was shown that for the FedWeb12 dataset \citep{Nguyen2012}, the inter-assessor disagreement was much stronger than the intra-assessor disagreement. The PRM does not explicitly model the intra-assessor disagreement. However, if the assessors lack consistency with their own judgments, this will also increase the level of disagreement between assessors, which is captured by the PRM.

We first provide an overview of related work in \secref{sec:relatedwork}, focusing on assessment disagreement and graded relevance measures, before detailing the PRM in \secref{sec:PRM}. We then present the datasets (\secref{sec:setup}) that we will use to quantitatively study the model. \secref{sec:relweights} explains how gains are set in the PRM, \secref{sec:analysis} analyzes the PRM parameters, and \secref{sec:PRMeval} presents retrieval system evaluation results. We conclude the paper and provide ideas for future research in \secref{sec:conclusions}.


\section{Related Work}
\label{sec:relatedwork}
The Predicted Relevance Model (PRM) introduced in this paper is related to evaluation approaches that investigate the plurality between users and assessors as well as effectiveness measures that use graded relevance levels. This section presents related work on these two aspects.

\subsection{Modeling Disagreement on Relevance}
Differences in relevance assessments for a particular result can originate from the actual difference in opinion by assessors, or from an error, \eg due to ambiguous instructions. These phenomena are often jointly referred to as assessment disagreement. 

Early works \citep[\eg by][]{Harter1996,Voorhees2000,Sormunen2002} study the influence of assessor errors on information retrieval evaluation using binary relevance judgments, and conclude that assessment disagreement has only minor effects on search engine comparisons. Based on these works, one could question whether extended models of assessors, such as our PRM, can make a difference. However, \citet{Bailey2008} aggregate and compare these early works and demonstrate that assessors with a different task and domain expertise significantly affect search engine comparisons. 

Furthermore, \citet{Vakkari2004} study assessors\footnote{\citet{Vakkari2004} adopt the term `users' for the persons reassessing documents. In our terminology, such persons are referred to as assessors.} that reassess results judgments by TREC assessors in an interactive search scenario. They mainly observe disagreement between their assessors and the TREC assessors at a lower relevance level, with a better agreement on highly relevant results. The distribution of disagreement over the relevance levels does not impact the validity of our current work, \ie the PRM remains valid for disagreement on lower levels \citep[as in][]{Vakkari2004}, and on top levels \citep[as in][]{Demeester2014}.

\citet{Carterette2010} also show that assessor disagreement has an effect on system comparisons for effectiveness measures considering graded relevance levels. Their identification of several prototypical assessor types (\eg unenthusiastic, pessimistic, lazy) is particularly relevant for test collections that route away from trained and supervised judges towards poorly trained and autonomous judges, \eg in  crowd-sourcing contexts. They show how different types of assessor errors affect evaluation measures, and propose strategies to compensate for such errors, \eg by reassessing certain results. \citet{Turpin2009} study the differences of using assessments that include summaries instead of only full documents on system comparison. They find that system effectiveness depends on the information that assessors have in order to make their decisions. \citet{Al-Harbi2014} investigate the difference in annotation behavior between so-called primary assessors, who create and judge test topics, and secondary assessors, who are paid to judge existing topics based on given query descriptions, and are less certain in their judgments. 

Our work differs from these contributions because the PRM does not assume a single truth label for each document and query, and uses assessments solely as predictions of the relevance for an unknown future user. Also, the source of the disagreement does not impact the way it is modeled in the PRM and used for predicting relevance towards a random user.
In an ideal scenario, assessment disagreement such as random annotation mistakes, which are complementary to the actual disagreement of assessors, should first be filtered from the assessments. However, the PRM is also suited to cope with these errors, and by modeling the resulting uncertainty on the assigned assessments ensures for a robust evaluation setting.  

An alternative to the classical assessment of isolated results, are preference judgments, which lead to higher agreement levels \citep{Carterette2008}. \citet{Kazai2013} examine the relationship between assessor disagreement and click based measures, which more directly reflects web users. Trained assessors appear to have higher inter-assessor agreement and are more likely to agree with clicks. Their results suggest that pairwise judgments lead to more awareness of the possible intent, and therefore lowers disagreement. The approach of preference judgments is not further investigated in the current work.

\citet{Hosseini2012} concurrently model the relevance of documents and the accuracy of individual assessors, based on multiple labels per document. Compared to this work, our PRM instead uses a limited set of documents with double judgments to model the ability of average assessors in predicting the average relevance according to a well-defined notion of user relevance, and is applicable to documents with single judgments. 

Our PRM model mainly focuses on dealing with potential disagreement of users/assessors on the \emph{relevance} of individual results. Besides in their judgment of relevance, users may also differ in the actions they take when browsing \emph{ranked} results. For instance, \citet{Carterette2012a} use click logs to compute posterior distributions for probabilistic models of user interactions and show that different ``types'' of user behavior exist, each of which may lead to a potentially different search system evaluation ranking. 
Metrics for that evaluation that are based on such more complex user models (which we leave out of scope for this paper on relevance disagreement) include rank-biased precision (RBP) \citep{Moffat2008}, expected reciprocal rank (ERR) \citep{Chapelle2009}, expected browser utility (EBU) \citep{Yilmaz2010}, and time-calibrated measures \citep{Smucker2012}. Such metrics mainly aim at appropriately accounting for the impact of the \emph{rank} a result is placed at. The current paper rather focuses on setting the appropriate weight of a result depending on its \emph{relevance}, as in graded relevance effectiveness measures, as discussed next. 
We note that \citet{Smucker2012} also integrate into their model the probability that a user considers a result relevant, by clicking or saving it, given that a NIST assessor judged it as relevant. In that aspect, there is a connection to the PRM approach.

\subsection{Graded Relevance Effectiveness Measures}
Graded relevance effectiveness measures allow assessors to use more than binary relevance labels. According to \citet{Kanoulas2009}, there are two main challenges with this class of effectiveness measures: (i) to set the gain for each relevance label, and (ii) to define the discount of this gain according to the rank of the document. \citet{Jarvelin2002} propose the popular normalized discounted cumulative gain (nDCG) measure, which uses a heuristic to set the relevance level gains and a logarithmic discount per rank. However, \citet{Kanoulas2009} find that gains should be set according to a user model in order to ensure that the measure reflects real users. \citet{Zhou2014} propose to learn suitable gain and discount functions based on assessor preferences of rankings. 
\citet{Sakai2007} compares 14 graded relevance measures with 10 traditional binary measures, and concludes that average nDCG at rank $k$ (nDCG@k) is among the best effectiveness measures for graded relevance in terms of stability, sensitivity, and resemblance of system rankings, and is fairly robust
with respect to the choice of gain values.

\citet{Chapelle2009} propose the Expected Reciprocal Rank (ERR)  effectiveness measure, which measures the inverse expected effort required for a user to satisfy their information need, and assumes the knowledge of the probabilities $R_i$ that the user is satisfied with document $i$. The discount function is therefore based on a user model but the work does not specify a user model to set $R_i$. This is achieved by the effectiveness measure GAP by \citet{Robertson2010} where each individual user is imagined to have a threshold label, above which they consider documents relevant. GAP then determines gains based on the distribution of users over these thresholds. However, empirically determining threshold distributions is hard and may depend on the query. Like GAP, our PRM also considers users that actually have a binary notion of relevance. Unlike GAP, the PRM employs the differences in the prediction of assessors to set gains. As it is often simpler to observe assessors than users, setting these gains is easier and requires less data. Furthermore, although not studied in this paper, it seems plausible that the parameters of our PRM can be used to arrive at the probabilities $R_i$ of the ERR measure.

\citet{Voorhees2001} studies the difference in effectiveness between using all relevant results and only highly relevant results. She finds that graded relevance measures are unstable due to the low number of highly relevant documents. Our PRM model improves the stability of graded relevance effectiveness measures by smoothing the judgment of an assessor with the possibility that users disagree with this judgment. For example, it takes into account the probability that a random user may consider a result top despite a judgment below the top level.

The model presented in this paper leads to estimates of the gains for the various relevance levels according to the probability that a random user consider results assessed with these levels as relevant. Our model hence assumes a binary notion of user relevance. 
However, other choices are possible. \citet{Kekalainen2005} and \citet{Voorhees2001} propose weighting of relevance grades based on the (speculated or heuristic) relative importance of the relevance levels to the users. Compared to binary relevance, this leads to a more complex notion of user relevance, and hence a more flexible evaluation scenario. The current work could be combined with these approaches. When adopting a more general (non-binary) model of user relevance, the probability of agreement with this user model, given an assessor judgment and based on the average disagreement, could be used to properly adapt the relevance gains. This however falls out of scope for the current paper.


\section{The Predicted Relevance Model}\label{sec:PRM}

The Predicted Relevance Model (PRM) presented in this section formalizes and extends the ideas from our experimental investigation of user disagreement in \citet{Demeester2014}, where also the original User Disagreement Model (UDM) was put forward. %
In the following we first formally define the PRM (\secref{subsec:defPRM}). We then describe a first application of the PRM in counting relevant results (\secref{subsec:countrel}), which allows transforming binary evaluation measures into graded measures based on the probability of binary relevance for an average user (\secref{subsubsec:binary}) and leads to an interpretation of using the nDCG measure with PRM-based gains (\secref{subsubsec:graded}). Finally, the differences and advantages of the PRM with respect to the original UDM formulation are discussed (\secref{subsec:UDM}).


\subsection{Definition of the PRM}\label{subsec:defPRM}

In this section, we provide the definitions behind the PRM, each followed by the key ideas on
\inlinelist{(\roman*)}{
\item the distinction and link between users and assessors,
\item the quantification of disagreement,
\item the conditions for the validity of the PRM,
\item the construction of relevance gain values.
}

\newtheorem{define}{Definition}

\begin{define}\label{def:users}
The considered user population of the IR system or search engine under evaluation, consists of individual users for whom a result is either relevant ($R$), or non-relevant to a query. 
\end{define}

\begin{define}\label{def:assessors}
The assessors are part of the evaluation setup, and assign relevance labels to results, according to well-described graded (or categorical) assessment levels, indexed by $i=0,\ldots,T$. The lowest level $i=0$ represents non-relevance, and the highest level $i=T$ is defined as top relevance. 
\end{define}

\paragraph{Users vs.\ assessors\\*}
The distinction between \emph{users} and \emph{assessors} is essential to the PRM. 
The \emph{user} model, on the one hand, corresponds to the classical binary notion of relevance for each individual user. Different users may have different opinions on the relevance of the same result.
The \emph{assessors}, on the other hand, are an essential part of the setup to evaluate retrieval systems. They assign different relevance grades according to how useful they predict a particular result to be to the users. A description of these relevance grades is part of the evaluation setup, and identical for all assessors. The assessor model corresponds with the typical scenario of graded relevance assessments for IR evaluation.

\paragraph{Intuition of the PRM\\*}
As will be described in the following sections, this setup allows evaluating how capable a system is in returning results relevant to a random user.
The intuition behind it can be summarized as follows.
A given result will be considered relevant by one user, while another might find it not relevant. The task of the assessor in an evaluation setup, usually implicitly amounts to try and assess how likely it is that a \emph{random} user would consider the given result relevant. This is typically done using graded relevance levels. 
How informative the assessments are, depends on how well the assessors are able to put themselves in the position of a user, and on the average user disagreement.
This intuition leads to the definition of parameters that quantify disagreement on relevance.

\begin{define}\label{def:p}
$p_{R\vert i} \defis $ the probability that a random user would consider a particular result relevant ($R$), given the knowledge of an independent assessor judgment with level $i$. 
\end{define}

\paragraph{Disagreement parameters\\*}
With Definition~\ref{def:p}, we model a particular result's relevance to a random user as a Bernoulli distributed variable. In fact, we model the user relevance of \emph{any} result for which an assessment with level $i$ was observed, as a Bernoulli variable with success rate $p_{R\vert i}$. 
%
The parameters $p_{R\vert i}$ are called the disagreement parameters, as they are subject to the disagreement between an assessor and a user. 

\paragraph{Assessors judgments for predicting user relevance\\*}
In a practical evaluation setup, the disagreement between user and assessor will be modeled from observations between different assessors, as these are the only ones observed. 
Consequently, the model only allows making claims towards the user population if the assessors are capable of putting themselves in the position of the user, at least on average over their assessments. 
%
%
%
This condition is not obvious in practice. Primary assessors are more likely able to judge results from the perspective of users, whereas secondary assessors are known to make more uncertain decisions~\citep{Al-Harbi2014}. The observed disagreement among secondary judges, or between primary and secondary judges, may therefore not reflect the disagreement with respect to users. Specific details on the assessors in our experimental setup will be given in \secref{sec:setup}.

\paragraph{Linking user model and assessor model\\*}
Even if the assessors can put themselves in the position of typical users, another condition needs to be fulfilled, in order to move from modeling disagreement between assessors in terms of assessment levels, to modeling binary relevance for users. We need to be able to map the assessment levels to the binary notion of user relevance. How this is done, depends on the goal of the evaluation, and the nature of the assessment levels. Although the PRM remains valid for any set of categorical relevance levels, for this paper we make the simplifying assumption that they are graded, and can be indexed from the lowest to the highest level $i=0,\ldots,T$, as in Definition~\ref{def:assessors}. In that case a logical choice to define the user model is by means of a threshold $\theta$ on the assessment levels. For levels on or above this threshold ($i\geq\theta$) the user model assumes relevance, and non-relevance below the threshold ($i<\theta$). We illustrate this with two examples.

In a web search evaluation scenario, the goal may be to evaluate a system's capability of retrieving top relevant results, for example for highly precision-oriented applications. In that case, we assume that users are only satisfied with top results. This means the threshold for user relevance is at the top graded relevance level ($\theta=T$). This was the choice made for the initial introduction of the UDM by~\citet{Demeester2014}. 

In a more lenient evaluation scenario, typical for recall-oriented applications, the user relevance threshold could be chosen just above non-relevance ($\theta=1$), to indicate that users are satisfied with any at least marginally relevant result, or to test a search engine's capability of filtering out non-relevant results.
Defining user relevance in such a way allows adapting the evaluation strategy for different applications or types of users. 

\paragraph{Asymptotic case without disagreement\\*}
In the asymptotic case of a perfectly controlled environment with deterministic annotation rules, and without any disagreement among assessors or users, we would find $p_{R\vert i\geq\theta}=1$ and $p_{R\vert i<\theta}=0$. Evaluation based on the PRM would boil down to classical binary evaluation at threshold $\theta$. 

\paragraph{Extension towards multiple random users\\*}
Definition~\ref{def:p} can be extended by considering the binomial distribution of relevance over multiple users, instead of the Bernoulli distributed relevance of a single user. For example, based on $p_{R\vert i}$, the probability can be calculated that at least $M$ out of $N$ random users would consider the result assessed with level $i$ as relevant, as shown by \cite{Demeester2014}. This allows rescaling the disagreement parameters in a consistent way, with a probabilistic interpretation, in order to adapt the evaluation setup towards a stricter or more lenient interpretation of relevance. This will not be pursued any further in the current work.


\paragraph{Modeling assessor behavior\\*}
A final point of discussion is on situations in which the assessors are not able to imitate the users' notion of relevance, 
and thus the PRM cannot make claims on the user population, as indicated before.
Even then, using the PRM has clear advantages with respect to heuristics, although the evaluation scenario would only model the assessor behavior, not the actual user population. 
For example, in the particular case of noisy crowd-sourced relevance judgments, it is doubtful that the assessors have a good understanding of what the users are like. 
However, the goal is still to use these assessments to evaluate IR systems. Given the large assessor disagreement, the probability $p_{R\vert i<T}$ would be quite high, or $p_{R\vert T}$ rather low, as confirmed by \citet{Demeester2014}. An evaluation based on a heuristic choice of relevance gains, such as gains exponential in the relevance grade $i$, may rely too strongly on the top judgments, and lead to a questionable robustness with respect to the choice of assessors. 
The PRM gains are adapted to the disagreement, and prevent an incorrect resolution between the systems under evaluation if the assessor disagreement does not allow it. This is in line with the work from \citet{Smucker2012}, who argue that metrics which fail to model user variance overestimate the effect size of differences between retrieval systems. 
For example, consider the extreme case that relevance grades are randomly assigned. When comparing retrieval systems based on these assessments, no valid conclusions can be made. A traditional evaluation setup would incorrectly favor IR systems that highly rank top judged documents, especially if based on a limited number of test topics. According to the PRM, however, no difference between any of these systems would be detected, because the gains for all relevance grades would have equal values.

\paragraph{Applicability of the PRM\\*}
We conclude by saying that the PRM is widely applicable, taking into account disagreement between assessors. Whether the results allow making conclusions about the user population, or only represent the assessors, depends on how well assessors are able to judge from the users' perspective. This holds in general when evaluating IR systems based on assessor judgments, just like the assumption that the judged search results and test topics are representative for how the systems will be used in practice.


\subsection{Counting Relevant Results}\label{subsec:countrel}


Before showing how the PRM can be used to set relevance weights in existing evaluation measures, we consider the task of counting the number of relevant results $N_R$ in a set of $N$ results, each with an associated relevance assessment. Let $n_i$ indicate the number of results assessed with level $i$ (with $\sum_i n_i = N$).
The link between the binary user model and the assessment grades is defined by a threshold $\theta$, as described in the previous section. 
If we neglect any disagreement between assessors or users, and purely estimate the number of relevant results from the assessments, we find
\e 
N_{R}^{\text{bin}} = \sum_{i\geq \theta} n_i,\label{eq:count_binary}
\f
in which the superscript `bin' indicates the binary model based on the assessments alone.
Taking into account the disagreement, the PRM leads to 
\e 
N_{R}^{\text{PRM}} = \sum_{i=0}^T n_i \> p_{R\vert i}.\label{eq:count_PRM}
\f
Equation~(\ref{eq:count_PRM}) is the summation for each relevance grade $i$ of the expected values $n_i \> p_{R\vert i}$ of the binomially distributed number of relevant results, given an observed assessment with level $i$, in $n_i$ trials. This leads to the interpretation of $N_{R}^{\text{PRM}}$ as the total expected number of relevant results in the results set for a random user.
In the following section we show how this result can be used to interpret evaluation measures that make use of the PRM, and in \secref{subsec:countFW13}, we will give an experimental illustration.

\subsection{The PRM and Evaluation Measures}\label{subsec:PRMeval}

This section describes how the PRM can be applied to binary evaluation measures that are based on counts of relevant results (\secref{subsubsec:binary}), and how the nDCG measure can be interpreted from the PRM perspective (\secref{subsubsec:graded}).

\subsubsection{Binary Evaluation Measures}\label{subsubsec:binary}
There are a number of established binary evaluation measures that rely on the number of returned relevant results. For such measures, the assessed number of relevant results in a traditional binary setting can be replaced by the expected number of relevant results according to the PRM. For measures that are linear in the number of relevant results, this leads to the expected value of that binary measure for a random user, as opposed to the value for the assessor alone. For example, the expected precision at rank $N$ based on the PRM would be $N_{R}^{\text{PRM}}/N$, with eq.~(\ref{eq:count_PRM}). This allows transforming a binary evaluation measure effectively into a graded measure, still measuring binary relevance for users, but whereby the weights of the relevance grades represent the uncertainty on the assessors' judgments with respect to user preferences.

\subsubsection{Graded Evaluation Measures}\label{subsubsec:graded}
A similar reasoning is also possible for graded relevance measures. We will discuss the case of the normalized discounted cumulative gain (nDCG) measure, given its popularity. We will thus also use nDCG for our experimental results in \secref{sec:PRMeval}. The application of the PRM to other measures 
is left open for future research.

Given a ranked results list, the nDCG measure incorporates the relevance of the result at rank $r$ by means of the gain $g(i(r))$ which is a function of the relevance level $i$ of that result.  
The cumulative gain at rank $k$ (CG@k) is defined as %
the sum of the gains for each result up to that rank. Typical gain values used in literature are the exponential gain $(2^{i(r)}-1)$ or the linear gain $i(r)$. 
Assuming that results at higher ranks are less likely to be reached by the user, the discounting factors $c(r)$ are introduced, leading to the discounted cumulative gain at rank k, similar to \citet{Zhou2014}, as
\e
\text{DCG@k} = \sum_{r=1}^k c(r) \> g(i(r)).\label{eq:dcg}
\f
The discount factors used most often in literature are the logarithmic discount $c(r)=1/\log(r+1)$, in which the gain a user obtains by moving down a ranked list drops less sharply than with the Zipfian discount $c(r)=1/r$~\citep{Kanoulas2009}. 
The nDCG@k measure is obtained by normalizing DCG@k calculated from the ranked list of retrieved results, by the ideal DCG@k when based on a perfect ranking, \ie according to decreasing relevance levels. 

We propose to calculate the nDCG@k measure with the PRM disagreement parameters as gains, $g^{\text{PRM}}(i)=p_{R\vert i}$, in order to model the relevance towards an average user.
The choice of discount factors remains open, as the PRM is not suited to model the rank-dependence of relevance in a results list. For our experiments, we use the logarithmic discount function.
Our proposal for using the disagreement parameters as relevance gains can be motivated as follows, in a similar way as in \secref{subsubsec:binary}, \ie by considering the binary relevance perspective for a random user. 

We assume the binary notion of user relevance introduced in \secref{subsec:defPRM}, based on a threshold $\theta$ on the relevance grades. The corresponding binary gain values can be defined as $g^{\text{bin}}(i)=1$ if $i\geq\theta$, and $g^{\text{bin}}(i)=0$ otherwise. The cumulative gain CG@k based on $g^{\text{bin}}(i)$ can be interpreted as the number of relevant results among the top $k$ retrieved results purely based on the assessor, ignoring any disagreement. The binary DCG@k, according to eq.~\ref{eq:dcg} but based on $g^{\text{bin}}(i)$, reduces to summing the discount factors of those ranks ($r\leq k$) with a result on or above the threshold ($i(r)\geq\theta$). 
The normalization factor for the binary nDCG@k is calculated as the binary DCG@k for the ideal ranking that places all results with grade $i\geq\theta$ before the others.

Using the PRM gains $g^{\text{PRM}}(i)$ leads to the interpretation of the resulting CG@k as the expected number of relevant results up to rank $k$, and of the resulting DCG@k as the expected value of the binary DCG@k, for a random user. The ideal ranking needed for the normalization in nDCG@k is based on decreasing relevance gains, in other words, based on the decreasing probability of relevance to a random user, given the assessor label. 

With this approach, no ad-hoc quantification of the relevance level gains is needed. The relevance gains emerge naturally as the PRM disagreement parameters when calculating the expected value of the binary DCG measure for a random user.


\subsection{Advantages of the PRM vs.\ the UDM}\label{subsec:UDM}

This section explains the differences between the PRM and the UDM, focusing on the differences between the respective user models. 

\paragraph{The UDM user model\\*}
The UDM introduced by~\citet{Demeester2014} is based on a different user model, compared to the PRM presented in the current paper. 
The UDM relevance weights correspond to the probability that at least either a random assessor, or the observed one, would consider a particular result a top result, given the relevance level assigned by the latter\footnote{The UDM was actually defined based on the probability that at least $M$ out of $N$ assessors, including the observed one, assign the top level. However, based on the binomial distribution, this is a straightforward extension from the case of $M=1$ and $N=2$, which is described here and corresponds best to the PRM formulation.}. 
As a result, the weight assigned to a result assessed as top relevant becomes one, and the weight for levels assessed below the top level ($i<T$) corresponds to the probability $p_{T\vert i}$. 
The sum of the UDM relevance weights over a set of results is the expected number of results based on the UDM user model. 
This corresponds to the expected number of results with at least one top level score by the observed assessor or a random one, and is obtained by adding up the actual number of results assessed with the top level, with fractional counts $p_{T\vert i<T}$ for lower rated results.

\paragraph{The PRM user model\\*}
The parameters of the PRM are based on a simpler and more intuitive user model, whereby we predict the relevance for a random user, again based only on the knowledge of an assessment level. By modeling a random user and leaving out the assessor, instead of selectively accepting those assessments with the highest level as true in the UDM, we do not enforce the top level relevance weight to be one, as in the UDM. 
The simplicity of the PRM user model leads to the interpretation of the summed disagreement parameters over a set of results as the \emph{expected} number of relevant results for a random user.

\paragraph{Counter-intuitive results with the UDM\\*}
Although sound by itself, the original UDM user model is less intuitive than the new PRM user model, and may lead to counter-intuitive results in special situations. 
For example, consider the case where the top relevance level ($T$) and the second highest relevance level ($T-1$) are conceptually very close to one another (e.g., $T$ defined as `Top result', and $T-1$ as `Excellent match', such that the distinction between both levels becomes really difficult for assessors). The confusion between these levels would yield both $p_{T\vert T-1}\approx 0.5$, and $p_{T\vert T}\approx 0.5$. Intuitively, both relevance levels could be considered top levels, and should therefore have similar weights. While the UDM assigns the weight for the official top relevance level $T$ as 1, and approximately 0.5 for the other effective top level, the PRM would assign equal weights to both levels, following the intuition outlined above.

\paragraph{Link between user and assessor model\\*}
A further difference between the PRM and the UDM, is the link between the user and assessor model. Although the distinction between both models was made less explicit by \citet{Demeester2014} than in the current work, the UDM assumes that users are only satisfied with top results. The PRM is formulated in a more general way: relevance between users is defined separately from the assessment levels. As explained, it is convenient in practice if the various assessment levels can be mapped to the binary notion of user relevance. To this end, multiple choices are possible, with various interpretations of the evaluation scenario.

\paragraph{Gains for non-relevant results\\*}
In the UDM, the gain for the lowest relevance level ($i=0$) was defined as zero, whereas the PRM gain of non-relevant results is the possibly non-zero value of $p_{R\vert 0}$. Stating that results considered non-relevant by the assessor should have no contribution to evaluation metrics, as in the UDM, is convenient and in line with traditional evaluation strategies. However, we do not want to exclude situations where a random user might consider such a result relevant. In the PRM case, user relevance is not limited to the top assessment level as in the UDM. For example, if user relevance is captured by any assessment level above the lowest level (\ie with threshold $\theta=1$), confusion between user relevance $R$ and the lowest assessment level becomes more likely, and can no longer be ignored in general. 

In some cases, however, ignoring the contribution of $p_{R\vert 0}$ is allowed, which allows significantly reducing the additional annotation effort for estimating the disagreement parameters (see \secref{subsubsec:general}). 
Also, when the disagreement parameters are meant to represent the whole collection, e.g., obtained by randomly selecting documents for annotation, it may be convenient to explicitly set $p_{R\vert 0}$  to zero, as in the UDM. The large majority of documents are most likely completely non-relevant to a given query, and should therefore not contribute to the total relevance. 
Low yet non-zero values of $p_{R\vert 0}$ in this setting may be due to annotation errors.
In practical scenarios, however, the disagreement parameters would often be estimated from a biased subset of the data, intended for evaluation purposes, \eg by pooling search results. In most such cases, non-zero values of $p_{R\vert 0}$ cannot be neglected.
More details on how the disagreement parameters depend on the data subset used for evaluation, are given in Section \secref{subsec:quality}.


\section{Datasets} \label{sec:setup}
Before venturing into a more detailed analysis and discussion of the practical application of the PRM, we present the two data sets that we use to support that discussion with quantitative experiments, highlighting the properties and behavior of the PRM. Both datasets contain a (sub)set of double graded relevance assessments, and are as such ideal for experiments with the PRM.

\subsection{TREC 2013 - Federated Web Search Track}\label{subsec:data:FedWeb13}
The first dataset used in this work comes from the TREC 2013 Federated Web Search Track (FedWeb13) \citep{Demeester2013TREC}. 
This track was created to stimulate research in federated search and the dataset contains the actual results of 157 real web search engines, including both the returned snippets and the actual pages of the top-10 results for each query. The 2013 edition of the track featured a resource selection and results merging task. The goal of the resource selection task was to rank the different resources on their predicted relevance to the test topics. In the results merging task, participants had to create a single ranked list over the results from all resources. 
Although initially a large set of 200 test topics was provided to the participants, the evaluation itself was based on the judgements for 50 test topics.

Students with different backgrounds were recruited to judge the relevance of the results, covering the fields of engineering, law, computer science, music, economics, and arts. From the initial set of test topics, the students were assigned topics of their choice, according to their expertise, which they then had to entirely annotate. Although the queries were not judged by those who initially created the queries, they themselves wrote narratives on which the judgments were based, from their own perspective. Because they selected their own queries and defined the information need, it is reasonable to see them as primary rather than secondary assessors (see \secref{subsec:defPRM}). 

The relevance of search results was graded on the following  levels: \RL{Non} (not relevant), \RL{Rel} (minimal relevance), \RL{HRel} (highly relevant), \RL{Key} (top relevance), and \RL{Nav} (navigational). For our experiments, we merged the few \RL{Nav} labels into the \RL{Key} category (this was also done for the official task evaluation, as the test topics were not navigational in nature). The dataset contains 34,010 results for the 50 test topics, for which both the page and the snippet were (independently) judged. In addition, a subset of double judgments was collected for a subset of the data (6,253 for the snippets and 7,027 for the pages). These double judgments were mostly chosen at random, also depending on the availability of assessors. Sometimes only a few (\eg the first three) results from a result list were judged twice, sometimes all 10. In total, 26 of the test topics contain double snippet judgments, and 24 topics have double page judgments. The assessors that provided the second set of judgments for a particular query did not create the query or narratives themselves, but again judged queries they themselves could have created, according to their interests. As a result, the user population towards which the PRM parameters will be tuned, consists of students whose information needs are mostly informational. 

Further information on the data and the relevance judgments can be found in the FedWeb13 overview paper \citep{Demeester2013TREC}.
All judgments were released by the track organizers in the `Fedweb Greatest Hits Collection' \citep{Demeester2015}. 

For the FedWeb13 system evaluation experiments described in \secref{sec:PRMeval}, the 18 submitted runs by 9 teams for the resource selection task are used, as well as the 15 runs by 6 teams for the results merging task. The evaluation measures are calculated with the trec-eval software\footnote{\url{http://trec.nist.gov/trec_eval/}}.


\subsection{NTCIR-10 2013 - Intent-2 Task}\label{subsec:data:INTENT-2}
The second dataset used in this paper contains the relevance judgments for the NTCIR-10 INTENT-2 Task, more in particular the Document Ranking Subtask for Chinese and Japanese data. In this task, the participants were asked to return a ranked list of search results. The test queries were in part navigational in nature, and in part informational. For the latter, the participants were required to diversify their results to cover different navigational intents. 
The results to be manually judged were selected by means of fing over the submitted runs, with a pool depth of 40.
For these, full double judgments are available, both for the Chinese test topics (22 navigational, and 75 informational ones), and for the Japanese topics (of which 28 are navigational and 67 informational). All judgments were done on a three-level scale, with levels \RL{0} (non-relevant), \RL{1} (medium), and \RL{2} (highly relevant), and by hired assessors. For the evaluation, these paired judgments were combined into a set of single 5-level gains. The resulting reference set of 5-level labels contains 22,552 explicit Chinese judgments, and 13,172 Japanese ones. In the current paper, we only consider the double three-level judgments.
More details can be found in the INTENT-2 overview paper by \citet{SakaiNTCIR2013}, and the overview paper at the first INTENT task at NTCIR-9, by \citet{SongNTCIR2011}, which provides additional details.

For the INTENT-2 evaluation experiments, the 12 submitted Chinese runs from 3 teams and the 8 submitted Japanese runs from 2 teams for the document reranking subtask are used, in combination with the NTCIREVAL toolkit\footnote{\url{http://research.nii.ac.jp/ntcir/tools/ntcireval-en.html}}.


\section{Practical Calculation of the Relevance Gains}\label{sec:relweights}

The following section describes a standard IR evaluation scenario for which the PRM method applies. 
A detailed description of how the disagreement parameters can be calculated, is given in \secref{subsec:estimatep}, first in general, then in practice for the FedWeb13 and INTENT-2 data.


\subsection{General Recipe}\label{subsec:generalrecipe}

Compared to an evaluation scenario with a common ad-hoc choice of relevance level gains, the PRM comes with an extra annotation cost: it relies on additional judgments on a subset of search results, which are necessary to estimate the degree of assessor disagreement. 
The steps of the PRM approach are:

\begin{enumerate}
\item
Gather a single set of graded relevance judgments for the test topics. 
\item\label{misc:step_divide}
Optionally: if the test queries can be naturally divided into homogeneous subsets (such as informational and navigational queries), the disagreement can be separately modeled for them, and the evaluation setup separated. To this end, perform Step~\ref{misc:step_estimatep} for each of these subsets individually.
\item\label{misc:step_estimatep}
Perform the following steps on the data:
\begin{enumerate}
	\item\label{misc:step_annotate}
	Gather a second set of judgments for a subset 
 	of the previously annotated search results, each by another assessor than for the original judgment.
	\item
	Estimate the disagreement parameters $p_{T\vert i}$ for all relevance levels $i$ (see \secref{subsec:estimatep}).
	\item Apply these as gains in suitable evaluation metrics (see \secref{subsec:PRMeval}).
\end{enumerate}
\end{enumerate}

The possibility mentioned in step~\ref{misc:step_divide} of dividing the data into more homogeneous subsets (e.g., according to different types of queries) has the advantage that a possibly different disagreement behavior is more accurately reflected in the different sets of relevance weights, as will be illustrated in \secref{subsec:estimatep}. It however requires sufficient double annotations for each of these subsets, which makes it more costly in return. 

Another important point pertains to the selection of a subset of search results to be annotated a second time in step~\ref{misc:step_annotate}, from which the parameters $p_{T\vert i}$ will be determined. The distribution of the results (in terms of general search result quality) and the required number of double judgments are discussed in \secref{subsec:quality} and \secref{subsec:amount}, respectively.


\subsection{Estimation of the Disagreement Parameters $p_{T\vert i}$}\label{subsec:estimatep}

\subsubsection{General Strategy}\label{subsubsec:general}
The discussion below covers the case where manual judgments are expensive, and at most two judgments from different assessors can be gathered for a subset of the test results. For the case of crowd-sourced Web search judgments, \citet{Demeester2014} show that the case of multiple judgments per result is approximated quite well by using only double judgments to estimate $p_{T\vert i}$. If more than two judgments per result are available, the  formulas proposed below to estimate $p_{T\vert i}$ as a ratio of occurrence frequencies can be extended.

The two different sets of annotations for the chosen results subset (see Section~\ref{subsec:generalrecipe}), are denoted as the set from user (or user group) $U_1$ and the one from user (group) $U_2$. $U_1$ and $U_2$ may represent actual groups of assessors, or correspond to an arbitrary separation of each double judgment into two groups, if the double judgments were provided by a single group of assessors. 

As explained in \secref{subsec:defPRM}, the PRM relies on the assessors' capability of estimating user relevance. In practice, the assessment levels are defined such that the binary notion of user relevance can be obtained directly from them, \eg based on a threshold $\theta$. To keep the notations simple, we write `relevant according to the binary user model' as $R$, denoting either of the levels on or above the threshold, or $i\geq\theta$.
Over the double judgments on all considered test topics, we define $N_{U_2=R, U_1=i}$ as the number of results judged above the binary relevance threshold by $U_2$ and with level $i$ by $U_1$, and $N_{U_1=i}$ as the number of results judged with assessment level $i$ by $U_1$.
We can estimate $p_{R\vert i}$ as
\e
p_{R\vert i} = \frac{N_{U_2=R, U_1=i}}{N_{U_1=i}}.\label{eq:naiveestimatep}
\f
If both groups of assessors independently judged the same pool of results, an alternative estimation is given by 
\e
p_{R\vert i} = \frac{N_{U_1=R, U_2=i} + N_{U_2=R, U_1=i}}{N_{U_2=i}+N_{U_1=i}}.\label{eq:estimatep}
\f

In order to make the estimation procedure more tangible, \tabref{tab:examplep} illustrates the use of eqs.~(\ref{eq:naiveestimatep}) and~(\ref{eq:estimatep}) for estimating disagreement parameters. In an artificial setting with 20 documents ($d_1$ to $d_{20}$), the 3-level graded relevance judgments by assessors $U_1$ and $U_2$ with respect to a query are listed, followed by the different estimates of disagreement parameters $p_{2\vert i}$ with respect to the top level $2$. Note that in reality the counts need to be higher, in order to obtain good estimates.

\begin{table}
\caption{Illustration of using eqs.~(\ref{eq:naiveestimatep}) and~(\ref{eq:estimatep}) for estimating disagreement parameters based on 20 pairs of 3-level judgments (with labels 0, 1, or 2) by assessors $U_1$ and $U_2$.} 
\label{tab:examplep} 
\begin{tabular}{lp{1mm}p{1mm}p{1mm}p{1mm}p{1mm}p{1mm}p{1mm}p{1mm}p{1mm}p{1mm}p{1mm}p{1mm}p{1mm}p{1mm}p{1mm}p{1mm}p{1mm}p{1mm}p{1mm}p{1mm}}
\hline\noalign{\smallskip}
\textbf{results}        &$d_1$&$d_2$&$d_3$&$d_4$&$d_5$&$d_6$&$d_7$&$d_8$&$d_9$&$d_{10}$&$d_{11}$&$d_{12}$&$d_{13}$&$d_{14}$&$d_{15}$&$d_{16}$&$d_{17}$&$d_{18}$&$d_{19}$&$d_{20}$\\
\textbf{ass. 
$U_1$} & 2   & 2   & 1   & 0   & 1   & 1   & 0   & 1   & 2   & 1   & 1   & 1   & 2   & 0   & 1   & 0   & 1   & 1   & 0   & 0   \\
\textbf{ass. 
$U_2$} & 1   & 2   & 1   & 1   & 2   & 0   & 0   & 1   & 2   & 2   & 0   & 2   & 1   & 0   & 1   & 2   & 0   & 1   & 0   & 0   \\
\hline\noalign{\smallskip}
\multicolumn{6}{l}{\textbf{eq. (\ref{eq:naiveestimatep})}} & \multicolumn{7}{l}{\textbf{eq. (\ref{eq:naiveestimatep})}} & \multicolumn{8}{l}{\textbf{eq. (\ref{eq:estimatep})}}\\
\multicolumn{6}{l}{$p_{U_2=2\vert U_1=2}=2/4$} & \multicolumn{7}{l}{$p_{U_1=2\vert U_2=2}=2/6$} & \multicolumn{8}{l}{$p_{2\vert 2}=4/10$}\\
\multicolumn{6}{l}{$p_{U_2=2\vert U_1=1}=3/10$} & \multicolumn{7}{l}{$p_{U_1=2\vert U_2=1}=2/7$} & \multicolumn{8}{l}{$p_{2\vert 1}=5/17$}\\
\multicolumn{6}{l}{$p_{U_2=2\vert U_1=0}=1/6$} & \multicolumn{7}{l}{$p_{U_1=2\vert U_2=0}=0/7$} & \multicolumn{8}{l}{$p_{2\vert 0}=1/13$}\\
\noalign{\smallskip}\hline
\end{tabular}
\end{table}

The PRM is based on the assumption that the average disagreement only depends on the observed relevance level, and is independent of the particular assessor. 
In reality, for the latter \citet{Carterette2010} have shown that assessors may actually differ in the proportion of documents they find relevant. For two such users $U_1$ and $U_2$, that would lead to a difference between $p_{U_1=R\vert U_2=i}$ and $p_{U_2=R\vert U_1=i}$, while \myeqref{eq:estimatep} takes into account a higher number of judgments and leads to an averaged estimate. 

If $U_1$ and $U_2$ each contain judgments from multiple assessors, \myeqref{eq:estimatep} is still more robust, but using \myeqref{eq:naiveestimatep} would be sufficient. In some situations, the amount of required double judgments can be strongly reduced. If during the second assessment round it becomes apparent that the estimate of parameter $p_{R\vert 0}$ is negligible, the extra judgments (by $U_2$) can be continued on a subset of only those indicated above non-relevance by $U_1$.
This is illustrated in \secref{subsubsec:FedWeb13_p}. 
Note that in this case, \myeqref{eq:estimatep} is no longer valid and the one-sided estimation \myeqref{eq:naiveestimatep} must be used, because the distribution of the relevance levels by $U_2$ no longer corresponds with the one from $U_1$. For example, a large fraction of level 1 judgments by $U_2$ would be missing (correlated with those indicated with level 0 by $U_1$), whereas most top level judgments would be present, such that the estimate in \myeqref{eq:estimatep} would be artificially high.

Alternative estimates for $p_{R\vert i}$ can be devised, e.g., with smoothing based on a Dirichlet prior, to deal with low numbers of occurrence of certain label combinations. This is left open for future research.

\subsubsection{FedWeb13 Disagreement Parameters}\label{subsubsec:FedWeb13_p}

\begin{table}
\caption{Estimated $p_{R\vert i}$ ($\pm$ 1 std.) for top relevance ($\theta=\RL{Key}$) on the FedWeb13 data, for pages and snippets, and using \myeqref{eq:estimatep} vs.\ \myeqref{eq:naiveestimatep}.}
\label{tab:pFedWeb13} 
\begin{tabular}{lcccc}
\hline\noalign{\smallskip}
\textbf{FedWeb13} & pages \eqref{eq:estimatep} & pages \eqref{eq:naiveestimatep} & snippets \eqref{eq:estimatep} & snippets \eqref{eq:naiveestimatep}\\
\noalign{\smallskip}\hline\noalign{\smallskip}
$\pRL{Key}{Key}$ & 0.53 $\pm$ 0.02 & 0.52 $\pm$ 0.02 & 0.47 $\pm$ 0.02 & 0.49 $\pm$ 0.03 \\
$\pRL{Key}{HRel}$ & 0.27 $\pm$ 0.01 & 0.28 $\pm$ 0.02 & 0.25 $\pm$ 0.01 & 0.26 $\pm$ 0.02  \\
$\pRL{Key}{Rel}$ & 0.04 $\pm$ 0.01 & 0.05 $\pm$ 0.01 & 0.08 $\pm$ 0.01 & 0.09 $\pm$ 0.01 \\
$\pRL{Key}{Non}$ & 0.01 $\pm$ 0.00 &  --               & 0.01 $\pm$ 0.00 &  -- \\
\noalign{\smallskip}\hline
\end{tabular}
\end{table}

\tabref{tab:pFedWeb13} shows the estimated disagreement parameters for the FedWeb13 data, both for pages and snippets, for the case that users are only satisfied with top results ($R=\RL{Key}$).  
There appears to be a substantial confusion between both highest levels (\RL{Key} and \RL{HRel}), and less so for the lower levels. The disagreement on different levels is similar for the full pages and for the snippets. The standard error on these estimates is shown as well, which is highest for \pRL{Key}{Key} (as the combination of two \RL{Key} judgments occurs the least), but remains within a few per cent. 

Note that the standard deviation $\sigma$ on the estimate of $p_{R\vert i}$ can be estimated as follows, given that $p_{R\vert i}$ is the success rate in a binomial distribution:
$$\sigma_{p_{R\vert i}} = \sqrt{\frac{N_N}{N_D} \left( 1 - \frac{N_N}{N_D}\right) \frac{1}{N_D}},$$
where 
$N_N$ and $N_D$ represent the numerator and denominator, respectively, of \myeqref{eq:estimatep} or \myeqref{eq:naiveestimatep}, depending on the estimation method.

The estimates based on \myeqref{eq:estimatep} are compared in \tabref{tab:pFedWeb13} with those based on the one-sided estimate \myeqref{eq:naiveestimatep}. For the latter, only rejudgments of results originally judged above \RL{Non} were used, such that \pRL{Key}{Non} could not be estimated. Neglecting the contribution from the lowest level is allowed in this case, due to the very low confusion with respect to the top level ($\pRL{Key}{Non}=0.01$). 
The results calculated with \myeqref{eq:estimatep} are slightly more robust (\ie they take into account more top judgments, and display a lower standard deviation). However, the differences are small, and the total number of double page (snippet) judgments used for the one-sided estimation amounts to only 19\% for pages and 23\% for snippets compared to the estimates with \myeqref{eq:estimatep}. 

\begin{table}
\caption{$p_{R\vert i}$ estimated from FedWeb13 page judgments, for different thresholds $\theta$ of binary user relevance $R$.}
\label{tab:pFedWeb13_R} 
\begin{tabular}{lccc}
\hline\noalign{\smallskip}
\textbf{FedWeb13} & $ \theta=\RL{Key}$ & $ \theta=\RL{HRel}$ & $ \theta=\RL{Rel}$ \\
\noalign{\smallskip}\hline\noalign{\smallskip}
$p_{R\vert\RL{\scriptsize{Key}}}$ & 0.53 & 0.87 & 0.93\\
$p_{R\vert\RL{\scriptsize{HRel}}}$ & 0.27 & 0.65 & 0.88\\
$p_{R\vert\RL{\scriptsize{Rel}}}$ & 0.04 & 0.22 & 0.46\\
$p_{R\vert\RL{\scriptsize{Non}}}$ & 0.01 & 0.02 & 0.08\\
\noalign{\smallskip}\hline
\end{tabular}
\end{table}

\tabref{tab:pFedWeb13_R} provides the disagreement parameters for the FedWeb13 page judgments, for three different choices of the threshold that defines binary user relevance as a function of the assessment levels. The left column shows results for binary relevance on the \RL{Key} level ($\theta=\RL{Key}$), the middle column assumes that users are satisfied with results they think satisfy the descriptions of either \RL{Key} or \RL{HRel} (with threshold $\theta=\RL{HRel}$), and the right column assumes that all levels above \RL{Non} are relevant to the user ($\theta=\RL{Rel}$). Relaxing the notion of user relevance leads to larger probabilities $p_{R\vert i}$. For example, where only 53\% of the users would consider a result assessed with the \RL{Key} label effectively a key result, 93\% would consider it at least marginally relevant. For the recall-oriented user scenario $\theta=\RL{Rel}$, an observed assessment with label \RL{HRel} is almost as likely as a \RL{Key} assessment to lead to relevance for a random user. For the precision-oriented approach $\theta=\RL{Key}$, results assessed as \RL{HRel} are only half as likely to satisfy a user as results assessed with the \RL{Key} label.
Also note that $p_{R\vert\RL{Non}}$ is very small for $\theta=\RL{Key}$, due to the limited confusion between the top and the lowest level, whereas it is larger for $\theta=\RL{Rel}$, due to the disagreement between the levels $\RL{Non}$ and $\RL{Rel}$.


\subsubsection{INTENT-2 Disagreement Parameters}\label{subsubsec:INTENT-2_p}

\begin{table*}
\caption{Estimated $p_{T\vert i}$ ($\pm$ 1 std.) on different query types (all, navigational, informational) for the INTENT-2 data.}
\label{tab:pNTCIR} 
\begin{tabular}{lccccc}
\hline\noalign{\smallskip}
\textbf{INTENT-2}& & all & nav. & inf. (all intents) & inf. (top intent)  \\
\noalign{\smallskip}\hline\noalign{\smallskip}
Japanese& $\pRL{2}{2}$    &    0.51 $\pm$ 0.01    &    0.77 $\pm$ 0.02    &    0.49 $\pm$ 0.01  &    0.54 $\pm$ 0.01 \\ 
        & $\pRL{2}{1}$    &    0.19 $\pm$ 0.00    &    0.17 $\pm$ 0.02    &    0.20 $\pm$ 0.01  &    0.24 $\pm$ 0.01 \\ 
        & $\pRL{2}{0}$    &    0.03 $\pm$ 0.00    &    0.03 $\pm$ 0.00    &    0.03 $\pm$ 0.00  &    0.05 $\pm$ 0.00 \\ 
\noalign{\smallskip}\hline\noalign{\smallskip} 
Chinese & $\pRL{2}{2}$    &    0.37  $\pm$ 0.01     &    0.07  $\pm$ 0.02     &    0.41 $\pm$ 0.02  &    0.29 $\pm$ 0.03 \\ 
        & $\pRL{2}{1}$    &    0.03 $\pm$ 0.00    &    0.04 $\pm$ 0.00    &    0.03 $\pm$ 0.00  &  0.04 $\pm$ 0.00 \\ 
        & $\pRL{2}{0}$    &    0.00 $\pm$ 0.00    &    0.01 $\pm$ 0.00    &    0.00 $\pm$ 0.00  &  0.00 $\pm$ 0.00 \\ 
\noalign{\smallskip}\hline
\end{tabular}
\end{table*}

\tabref{tab:pNTCIR} shows the estimated disagreement parameters $\pRL{2}{i}$ for the INTENT-2 data, estimated from all double 3-level judgments (column `all' in the table). When we consider all queries together, there seems to be a fair agreement on the top level. The amount of confusion between the highest and the middle level (\ie \pRL{2}{1}) is however much higher for the Japanese than for the Chinese data. This disagreement on the Japanese queries was already noticed by~\citet{SakaiNTCIR2013}, without giving any rationale behind it.

Two query types can be distinguished: navigational (22 Chinese and 28 Japanese queries) and informational (75 Chinese and 67 Japanese queries). 
The informational queries contribute more strongly to the combined results (column `all') than the navigational ones, because there are more of them, and they have multiple intents.
To investigate the influence of the navigational queries, we also calculated the disagreement parameters separately for the different query types, in agreement with step~\ref{misc:step_divide} of the general recipe (\secref{subsec:generalrecipe}). 
\tabref{tab:pNTCIR} illustrates clearly that these query types lead to a very different disagreement in both languages, such that making this distinction is justified and necessary. 

For the navigational queries (column `nav.'), there is a large difference in the level of agreement on the top results between both languages, which is very high for Japanese, and very low for Chinese.
For the latter, the fact that \pRL{2}{2} is so small, shows that there may be a problem with the relevance judgments, or at least with the assessors' interpretation of the top relevance level for a navigational query. 
 
For the informational queries, where multiple intents of the same query were separately judged, we considered different approaches to estimate the disagreement parameters. For the first approach (indicated as `all intents' in \tabref{tab:pNTCIR}), we considered each given (query, intent) pair as a different information need. The double judgments over all different intents and queries were taken together, and the parameters $\pRL{2}{2}$ were calculated with \myeqref{eq:estimatep}. 
Note that the judgments with intent label `0' (meaning none of the intents were judged relevant), were replaced by explicit separate judgments of non-relevance for each of the intents. 
The second approach (`top intent') is based on only the most probable intent for each query, given the intent probabilities \citep[as in][]{SakaiNTCIR2013}. The underlying idea is that the most probable intent for a query may lead to a different disagreement behavior than the average over all intents. Since even considering the top intents alone leads to enough judgments for confident estimates, Step~\ref{misc:step_divide} of the general recipe can be applied. For the Japanese data, the behavior remains the same, except for a small increase in the overall probability on a top judgment. For the Chinese data, there is an overall increase in the disagreement (lower \pRL{2}{2} and higher \pRL{2}{1}). 
In the remainder of the paper, the disagreement parameters as estimated from the top intents will be used. The reason is that for the evaluation part (\secref{sec:PRMeval}) the influence of the disagreement on the nDCG metric will be investigated, \ie to evaluate results on a single intent, as opposed to more advanced variations that account for intent diversity.


\section{Analysis of PRM Parameters}\label{sec:analysis}

In the following sections we take a closer look at some general properties and difficulties in applying the PRM, integrating experimental evidence immediately into the discussions. These issues include the dependence of the PRM parameters on the results quality (\secref{subsec:quality}), the choice of test topics (\secref{subsec:choicetopics}), and the number of double judgments (\secref{subsec:amount}).

\subsection{Sensitivity to Search Result Quality}\label{subsec:quality}

An important issue that may influence the final estimates for $p_{R\vert i}$ is the choice of the initial set of double annotations.
\citet{Webber2012} show that assessor disagreement on particular documents depends on the ranks at which these documents are retrieved by a set of retrieval systems, summarized into their `metarank': they model how the disagreement changes as a function of that metarank. This effect was also observed by \citet{Demeester2014} for the FedWeb12 dataset: if the $p_{\RL{Key}\vert i}$ parameters are estimated from high-quality results lists, they are larger than when estimated from average result lists. 
We now further explore this effect.

\begin{figure*}
\includegraphics{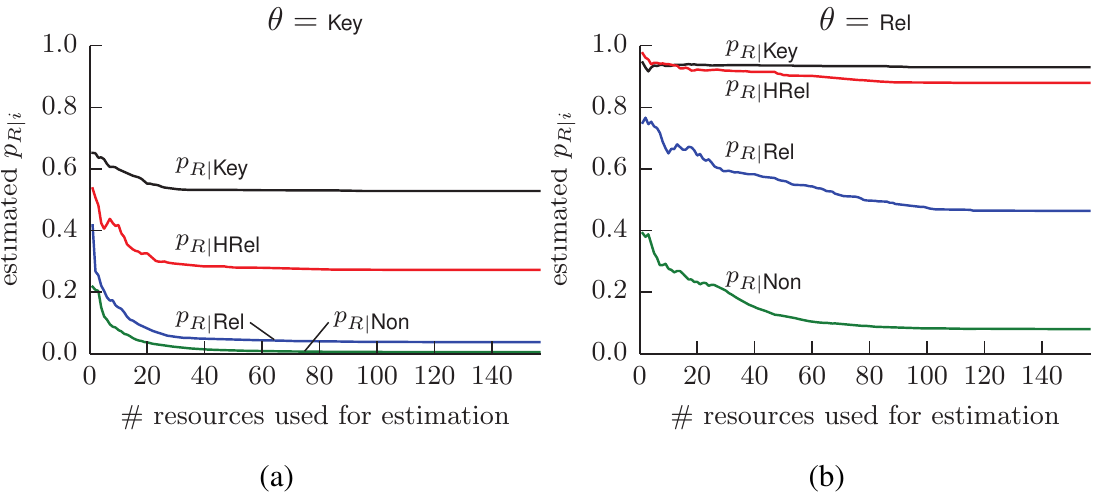}
\caption{Disagreement parameters for the FedWeb13 data, estimated using top-10 results from the top-$k$ ranked resources (in decreasing order of results relevance). We used double annotations of all results considered for the estimation.}
\label{fig:metarank}
\end{figure*}

\Figref{fig:metarank} visualizes the described phenomenon for the FedWeb13 data. 
For each query, we ordered all resources (\ie search engines) according to the descending number of \RL{Key} or \RL{HRel} results they returned (as measured by the single reference assessor for which full judgments are available). This leads to an ad-hoc ranking from high-quality to low-quality resources. The $p_{R\vert i}$ curves in \figref{fig:metarank} were obtained by gradually taking into account the top-10 results from more resources, starting from only the best resource, up to including them all. Two different scenarios for user relevance are shown: (a) with threshold $\theta=\RL{Key}$ corresponding with users that are only satisfied with top results, and (b) for $\theta=\RL{Rel}$, for users that are satisfied with any result at least marginally relevant. The asymptotic values, when all resources are taken into account, correspond to the values listed in \tabref{tab:pFedWeb13_R}. Note that in \secref{sec:PRMeval} we will further use these user scenarios, referring to them as the top relevance scenario ($\theta=\RL{Key}$) and the marginal relevance scenario ($\theta=\RL{Rel}$).

The disagreement parameters start high, when only high-quality resources are used, then decrease, and finally saturate as soon as the lower ranked resources contain no further results with the appropriate relevance levels to contribute. 
We would like to stress the fact that the judgments were done in a randomized order (within each query), where the assessors were not informed on the provenance (\ie resource) of the web page under assessment. This means that indeed the effect described by \citet{Webber2012} can be observed.
One possible explanation for the observed behavior is due to the limitations of using a small discrete set of relevance levels. 
Consider for example the observed behavior of $p_{R\vert \RL{Key}}$, for the relevance threshold $\theta=\RL{Key}$. 
Among all results assessed as \RL{Key}, we can imagine that some would receive an even higher relevance grade if it existed, with a correspondingly higher probability of an average user to consider it relevant. Such results considered more relevant than the average results judged as \RL{Key}, are more likely to come from the best resources, hence the elevated levels of $p_{R\vert \RL{Key}}$ if only these are taken into account. For the case $\theta=\RL{Rel}$, this effect on $p_{R\vert \RL{Key}}$ is very small, because apparently the explained variations among results indicated as \RL{Key} do not strongly influence the probability of a user to consider a result at least marginally relevant.

The take-away message of this discussion is the following. We have seen that the disagreement parameters may vary, depending on the set of results they are estimated from. Therefore, the main consideration for defining the set of double annotations to estimate $p_{R\vert i}$ from, is that it should be representative for the evaluation setting. For setups where the evaluation targets the higher-ranked results, those should be sampled from when gathering the double relevance judgments. 
For example, in the case of pool-based IR evaluation, if results up to a depth of 10 will be used for measuring system comparisons, a reasonable choice would be to gather the double annotations from a sample of the top-10 results by the systems under test.

For the experimental results shown in this paper, we have chosen to use the same disagreement parameters for the different evaluation settings, based on all available double judgments.


\subsection{Choice of Test Topics}\label{subsec:choicetopics}
\begin{figure*}
  \includegraphics{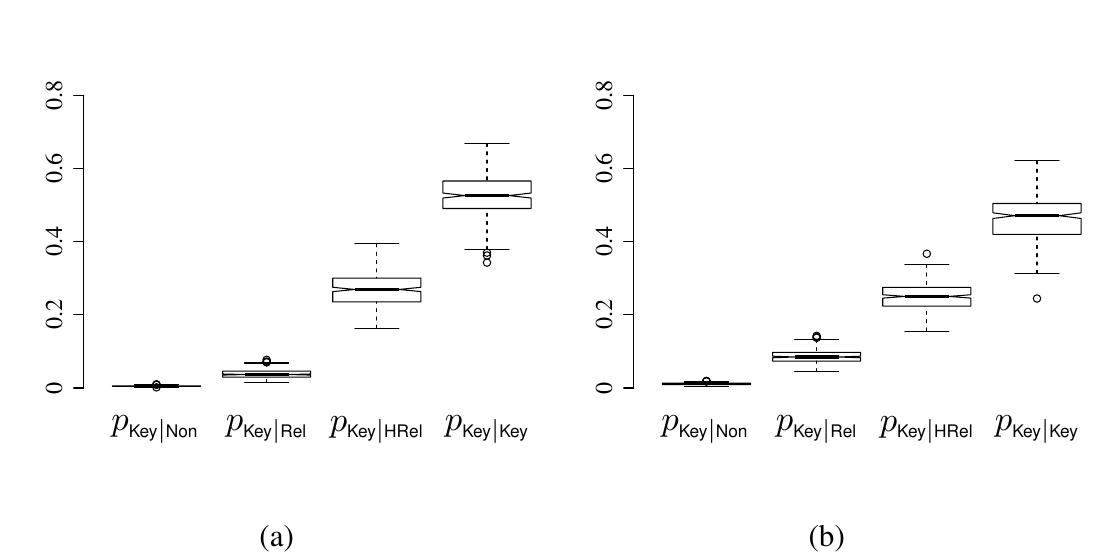}
  \caption{Boxplot of $p_{T\vert i}$ for different levels $i$, by bootstrapping the test topics for the FedWeb13 pages (a) and snippets (b).}
  \label{fig:boxplot_fedweb}
\end{figure*}

The disagreement parameters are calculated by aggregating double judgments over multiple test topics. However, the disagreement between assessors might depend on the particular topics, and yet the same PRM parameters are used for all topics. 
In particular, for the FedWeb13 data, approximately one out of five results was judged twice, distributed among half of the test topics (see \secref{subsec:data:FedWeb13}), and the relevance gains based on those are used to evaluate all 50 topics.
In order to visualize the dependence on the topics, we bootstrapped the different topics for which double annotations were performed, each time estimating the disagreement parameters based on the selected topics, in 300 bootstrap samples. For the calculations, all judgments for each of the topics were taken into account as many times as the topic was chosen for the particular bootstrap sample. \figref{fig:boxplot_fedweb} shows a boxplot of the result, both for snippets and pages, with a similar behavior. The largest variation occurs for the highest relevance levels, because their estimates are based on the fewest cases. For example for the pages, we find a standard deviation of 0.06 on the estimate of \pRL{Key}{Key} and 0.05 on \pRL{Key}{HRel}.
These values are higher than the corresponding standard deviations due to the total number of cases to estimate the disagreement parameters from, which are 0.02 and 0.01, respectively (see \tabref{tab:pFedWeb13}).
We conclude that the influence of the topics is noticeable, but does not invalidate the disagreement parameters because the variation is still limited. However, for using the PRM method, we recommend to gather incomplete sets of double judgments for a larger fraction of the test topics, as was done for the FedWeb13 data, rather than complete double judgments on a smaller number of topics, as previously done for the FedWeb12 data \citep[see][]{Demeester2014}. 

\subsection{Number of Double Judgments}\label{subsec:amount}

We now discuss the required number of double judgments. Given their extra annotation cost, ideally the number of double judgments should be kept to a minimum. The main requirement is that there are enough judgments to have a small enough uncertainty on the disagreement estimates. 
The allowed upper boundary of that uncertainty depends on the application. Yet, requiring that that the disagreement parameters for levels with a conceptually clear difference in relevance are well distinguishable, can be used as a sufficient condition for the number of double judgments.
Both for the FedWeb13 and INTENT-2 data, the standard deviations (shown in Tables~\ref{tab:pFedWeb13} and~\ref{tab:pNTCIR}) are small enough in that respect. The only exception is the vague distinction between the top and medium level for the Chinese navigational queries, due the very low top level agreement.

\begin{figure}
\includegraphics{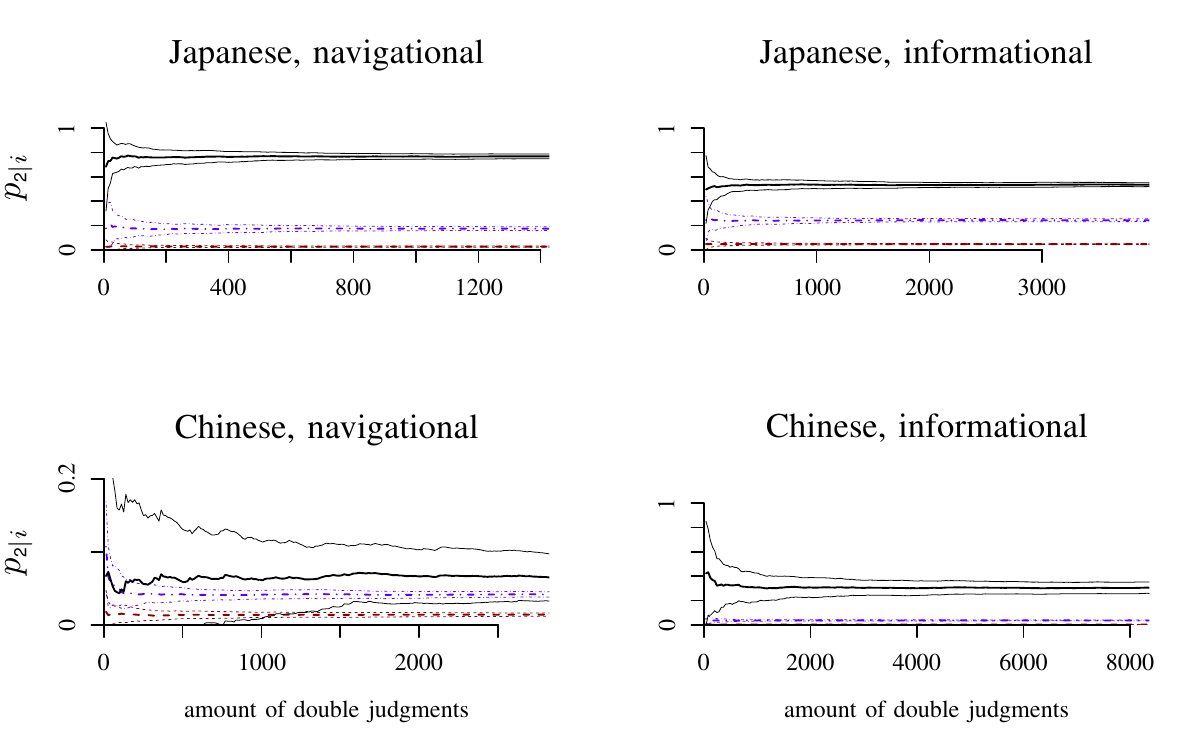}
\caption{Simulated $p_{\RL{\scriptsize{2}}\vert i}$ (mean $\pm$ 1 std.) on INTENT-2 judgments vs.\ the number of double judgments: \pRL{2}{2} (black full lines), \pRL{2}{1} (blue dash-dot lines), \pRL{2}{0} (red dashed lines).}
\label{fig:amountINTENT-2}
\end{figure}

To get an idea of the uncertainty on the estimates as a function of the required number of double judgments, we did the following bootstrap experiment on the INTENT-2 data, focusing on the scenario with paired judgments and \myeqref{eq:estimatep} to estimate $p_{\RL{\scriptsize{2}}\vert i}$. We simulated 50 annotation rounds by sampling the actual double judgments (with replacement), keeping track of the disagreement parameters for the growing set of simulated double judgments. The mean value and one standard deviation above and below it are shown in \figref{fig:amountINTENT-2}. 
Given the large amount of judgments (\ie full double judgments), the uncertainty on most of the estimates already becomes very small for a fraction of the judgments. For the Chinese navigational queries, however, the problems noted in \secref{subsubsec:INTENT-2_p} are confirmed. Even when the absolute uncertainty on \pRL{2}{1} and \pRL{2}{0} becomes small, they cannot be distinguished in terms of their disagreement with the top level. This makes the resulting parameters $p_{\RL{\scriptsize{2}}\vert i}$ less trustworthy, and any evaluation based only on these queries questionable.



\section{Application of the PRM for System Evaluations}\label{sec:PRMeval}

This section is devoted to the application of the PRM model to actual system evaluations, based on the INTENT-2 and FedWeb13 data. We will demonstrate the difference in counting the number of relevant results purely based on the assessor and as expected for a random user (\secref{subsec:countFW13}), demonstrate the robustness of evaluation with the PRM (\secref{subsec:robustness}), and investigate the behavior of system rankings based on PRM gains vs.\ heuristic gains (\secref{subsec:evaluation}).

\subsection{Counting Relevant Results}\label{subsec:countFW13}

As explained in \secref{subsec:countrel}, summing the disagreement parameters $p_{R\vert i}$ for each result in a result list, according to the assigned relevance level by the assessor, results in the expected number of relevant results according to a random user.
This allows making absolute conclusions about how well systems are able to return relevant results, whereas the ad hoc weighting (\eg linear or exponential) of relevance levels typically focuses on relative system comparisons, without a clear interpretation of the absolute value of the resulting metrics. 

Let us illustrate this with the results of the FedWeb13 Resource Selection (RS) task.
Participants were required to rank 157 online resources on their estimated capability of returning relevant results for a particular query. From 9 participating teams, results for 18 RS systems were submitted. 
In a typical federated search scenario, the results from the highest ranked resources per query are retrieved, merged into a single ranked list, and presented to the user. We only consider the top three resources per query. 
Given that per query, only the top-10 search results for each resource are available, the result sets that we evaluate 
for each system contain at most 30 results per query. 

\tabref{tab:numtop} shows the number of relevant results among the top three resources, averaged over 50 evaluation queries, together with the standard deviation on that average. The columns `PRM' show the number of relevant results a random user expects to find, estimated with eq.\ (\ref{eq:count_PRM}) according to the PRM model. 
The columns `binary' show the number of relevant results based on eq.\ (\ref{eq:count_binary}), purely based on single judgments, \ie ignoring the disagreement. 
Two different scenarios for user relevance are shown, the top relevance scenario ($\theta=\RL{Key}$), and the marginal relevance scenario ($\theta=\RL{Rel}$).

The results are shown for the 18 official runs\footnote{The TREC results are available at \url{http://trec.nist.gov/results/}.}, as well as two baselines by the organizers (\run{RS\_clueweb} and \run{RS\_querypools}), and an artificial RS system (\run{oracle}) that selects the three best possible resources per query.

\begin{table}
\caption{Number of relevant results among top 3 resources, for FedWeb13 Resource Selection runs (average over 50 test queries $\pm$ the st.\ dev.\ of the mean). PRM: expected number of relevant results for a random user; binary: number of relevant results by a single assessor. User scenarios: top relevance vs.\ marginal relevance.}
\label{tab:numtop} 
\begin{tabular}{lcccc}
\hline\noalign{\smallskip}
\textbf{run} & \multicolumn{2}{c}{top relevance  ($\theta=\RL{Key}$)} & \multicolumn{2}{c}{marginal relevance  ($\theta=\RL{Rel}$)}\\
 & PRM & binary & PRM & binary\\
\noalign{\smallskip}\hline\noalign{\smallskip}
\run{oracle                  } & $9.02\;\;(\pm 0.40)$ & $12.98\;\;(\pm 0.93)\;\;$ & $21.40\;\;(\pm 0.61)\;\;$ & $25.92\;\;(\pm 0.63)\;\;$ \\
\run{RS\_clueweb             } & $2.51\;\;(\pm 0.27)$ & $2.66\;\;(\pm 0.46)$ & $8.52\;\;(\pm 0.54)$ & $9.26\;\;(\pm 0.74)$ \\
\run{UiSSP                   } & $2.41\;\;(\pm 0.36)$ & $2.58\;\;(\pm 0.57)$ & $7.78\;\;(\pm 0.82)$ & $8.76\;\;(\pm 1.20)$ \\
\run{UiSP                    } & $2.27\;\;(\pm 0.38)$ & $2.52\;\;(\pm 0.63)$ & $7.22\;\;(\pm 0.84)$ & $8.00\;\;(\pm 1.17)$ \\
\run{utTailyNormM400         } & $2.05\;\;(\pm 0.37)$ & $2.20\;\;(\pm 0.54)$ & $6.65\;\;(\pm 0.86)$ & $7.24\;\;(\pm 1.14)$ \\
\run{utTailyM400             } & $1.94\;\;(\pm 0.37)$ & $2.06\;\;(\pm 0.54)$ & $6.32\;\;(\pm 0.86)$ & $6.74\;\;(\pm 1.13)$ \\
\run{UiSS                    } & $1.66\;\;(\pm 0.25)$ & $1.64\;\;(\pm 0.35)$ & $5.84\;\;(\pm 0.65)$ & $5.98\;\;(\pm 0.92)$ \\
\run{udelODRA                } & $1.63\;\;(\pm 0.30)$ & $1.68\;\;(\pm 0.48)$ & $5.64\;\;(\pm 0.73)$ & $6.00\;\;(\pm 1.01)$ \\
\run{udelFAVE                } & $1.63\;\;(\pm 0.29)$ & $1.62\;\;(\pm 0.43)$ & $5.78\;\;(\pm 0.73)$ & $6.14\;\;(\pm 1.01)$ \\
\run{UPDFW13mu               } & $1.54\;\;(\pm 0.32)$ & $1.52\;\;(\pm 0.44)$ & $5.18\;\;(\pm 0.79)$ & $5.36\;\;(\pm 1.06)$ \\
\run{iiitnaive01             } & $1.46\;\;(\pm 0.28)$ & $1.52\;\;(\pm 0.45)$ & $5.24\;\;(\pm 0.65)$ & $5.76\;\;(\pm 0.88)$ \\
\run{cwi13SniTI              } & $1.46\;\;(\pm 0.29)$ & $1.46\;\;(\pm 0.44)$ & $5.17\;\;(\pm 0.70)$ & $5.48\;\;(\pm 0.96)$ \\
\run{UPDFW13sh               } & $1.43\;\;(\pm 0.27)$ & $1.12\;\;(\pm 0.32)$ & $5.28\;\;(\pm 0.71)$ & $5.38\;\;(\pm 0.98)$ \\
\run{RS\_querypools          } & $1.25\;\;(\pm 0.19)$ & $1.06\;\;(\pm 0.34)$ & $5.33\;\;(\pm 0.44)$ & $6.18\;\;(\pm 0.80)$ \\
\run{cwi13ODPTI              } & $1.17\;\;(\pm 0.21)$ & $0.92\;\;(\pm 0.27)$ & $4.48\;\;(\pm 0.56)$ & $4.52\;\;(\pm 0.77)$ \\
\run{ECNUBM25                } & $0.91\;\;(\pm 0.22)$ & $1.16\;\;(\pm 0.38)$ & $3.04\;\;(\pm 0.64)$ & $2.62\;\;(\pm 0.56)$ \\
\run{cwi13ODPJac             } & $0.66\;\;(\pm 0.16)$ & $0.42\;\;(\pm 0.16)$ & $2.88\;\;(\pm 0.51)$ & $2.94\;\;(\pm 0.72)$ \\
\run{udelRSMIN               } & $0.61\;\;(\pm 0.21)$ & $0.78\;\;(\pm 0.33)$ & $2.28\;\;(\pm 0.48)$ & $1.94\;\;(\pm 0.62)$ \\
\run{incgqdv2                } & $0.55\;\;(\pm 0.13)$ & $0.48\;\;(\pm 0.21)$ & $2.17\;\;(\pm 0.35)$ & $2.08\;\;(\pm 0.46)$ \\
\run{incgqd                  } & $0.35\;\;(\pm 0.12)$ & $0.30\;\;(\pm 0.19)$ & $1.46\;\;(\pm 0.29)$ & $1.38\;\;(\pm 0.40)$ \\
\run{StanfordEIG10           } & $0.19\;\;(\pm 0.07)$ & $0.14\;\;(\pm 0.08)$ & $0.85\;\;(\pm 0.20)$ & $0.66\;\;(\pm 0.27)$ \\
\noalign{\smallskip}\hline
\end{tabular}
\end{table}

We can make a number of observations from these results. For both user scenarios, the PRM estimates of the average number of relevant results are more robust, given the lower standard errors, than the binary estimates. The system rankings between PRM and binary estimates are strongly correlated, although not the same: Kendall's tau is 0.93 for the top relevance scenario, and 0.89 for the marginal relevance scenario. 

We observe substantial differences in the absolute numbers of estimated relevant results, due to the difference between modeling disagreement (PRM) and accepting the assessors' judgments as ground truth (binary). Based on the disagreement parameters, these differences can be interpreted. For example in the user scenario of top relevance, two main effects play a role in the PRM results:
\begin{enumerate*}[label=(\arabic*)]
\item \label{it:disagree} The strong disagreement on the top level ($\pRL{Key}{Key}=0.53$) causes results assessed as \RL{Key} to contribute only half as much to the estimated number of top relevant results, compared to the binary estimate; 
\item \label{it:relvskey} Results only assessed as \RL{HRel} are considered \RL{Key} results by random users in about one out of four times ($\pRL{Key}{HRel}=0.27$).
\end{enumerate*}
For the oracle system, the top 3 resources contain 13 \RL{Key} results, purely based on the assessor, whereas a random user expects to find only 9 \RL{Key} results. This means effect \ref{it:disagree} is dominant. 
Some of the lower ranked systems have a higher PRM-based than binary estimate of the number of \RL{Key} results, for example the run \run{cwi130DPJac}. For such systems effect \ref{it:relvskey} dominates, and they are better at retrieving results assessed as \RL{HRel} than \RL{Key} results.

\subsection{Robustness of PRM-based Evaluation}\label{subsec:robustness}
A direct way to evaluate how well a system is capable of retrieving relevant documents, is by calculating effectiveness measures based on binary relevance: relevant results are rewarded, depending on the rank at which they are retrieved. 
Due to user disagreement on the top level, the evaluation scores and even score-based rankings between different systems may lack robustness. The PRM allows us to reward results based not only on the particular assessor's personal idea of user relevance, but on the expected relevance to a random user. Because the latter is estimated from the average disagreement between assessors, a PRM-based evaluation should lead to a more robust evaluation, with respect to the choice of assessors.  

This can be verified with the double set of 3-level INTENT-2 judgments, and the official runs submitted to the INTENT-2 Document Ranking Subtask. We consider user relevance at the highest assessment level $\theta=\RL{2}$: our evaluation reflects users that are only satisfied with top results.  
Each run is scored separately for the set of judgments from user $U_1$ and from user $U_2$. As an indicator of robustness, we consider Kendall's rank correlation coefficient $\tau$ between the resulting rankings of the runs, each based on one of the sets of assessments, \ie $U_1$ vs.\ $U_2$. As evaluation measure, we use nDCG@10, averaged over the test topics, and with a logarithmic discount function.
\tabref{tab:robustness} lists the results for the binary nDCG as introduced in \secref{subsubsec:graded} (column `binary'), for the PRM-based nDCG in which the disagreement parameters $p_{\RL{2}\vert i}$ are used as gains (column `PRM'), and with linear gains $g(i)=i$ (column `linear'). 

There is a clear difference between the Chinese and Japanese data: the order of the Japanese runs seems almost user-independent, whereas there is a strong mismatch for the Chinese data. This may in part be related to the limited amount of data: only 8 Japanese runs from 2 teams, and 12 Chinese from 3 teams. Another cause may be the organization of the assessments, since $U_1$ and $U_2$ actually contain judgments from multiple judges, but we cannot further investigate this effect, as the composition of $U_1$ and $U_2$ for both languages has not been made public. Yet, the main reason is the higher overlap on top judgments for the Japanese data, as opposed to the Chinese: we have \pRL{2}{2} = 0.77 and 0.54 for respectively the navigational and the informational queries in the Japanese data, while the Chinese has only \pRL{2}{2}~= 0.07, respectively 0.29.

\begin{table}
\caption{Kendall $\tau$ between system rankings based on different users for the INTENT-2 data, based on nDCG@10 with binary gains on top relevance, corresponding PRM gains, and linear gains.}
\label{tab:robustness} 
\begin{tabular}{lccc}
\hline\noalign{\smallskip}
\textbf{INTENT-2} & binary & PRM & linear\\
\noalign{\smallskip}\hline\noalign{\smallskip}
Japanese nav. & 0.86 & 0.86 & 0.64 \\
Japanese inf. & 0.84 & 0.93 & 1.00 \\
Chinese nav. & 0.12 & 0.43 & 0.47 \\
Chinese inf. & 0.06 & 0.27 & 0.72 \\
\noalign{\smallskip}\hline
\end{tabular}
\end{table}

The robustness of the evaluation based on $U_1$ and $U_2$ is finally also tested with linear gain values. In this case, the robustness also increases significantly with respect to the binary top evaluation. 
However, it is important to stress that there is an important conceptual difference between using the PRM and using linear gains. 
The choice of linear gains may be defensible in certain scenarios, but does not allow specifically testing the capabilities of a system in retrieving top relevant results, which both the top binary evaluation scenario and the associated PRM scenario do. 
For example, for the Chinese informational queries the linear gains lead to a higher robustness than the PRM, due to the stronger weighting of the medium levels. 
In this case the linear gain for the medium relevance level equals half the gain of the highly relevant results. What does this mean for the evaluation scenario? 
The PRM gain of level \RL{1}, \ie the chance that a random user would assign \RL{2} if the assessor said \RL{1}, is actually much lower than half the top level gain, or the corresponding chance if the assessor had said \RL{2}: a fraction 0.12. 
In other words, the linear gain of the medium level is too high to only account for disagreement on the top level. In this example, evaluation with linear gains not only rewards systems for retrieving top results, it also rewards them for their capability in retrieving medium results. Evaluation with linear gains is therefore not in line with the user model behind the binary and PRM gains, \ie user relevance for top results. 

A disadvantage of using fixed heuristic gains, is that interpretations as the one above are data-dependent. For example, in situations with very high disagreement, a linear gain might even not be high enough to compensate for the confusion of a particular level with the levels $i\geq\theta$. 
The PRM, in contrast, has an underlying evaluation scenario with a direct interpretation.


\subsection{Evaluation with PRM gains vs.\ Standard Gains}\label{subsec:evaluation}


\begin{table}
\caption{FedWeb13 Results Merging evaluation: Kendall $\tau$ between nDCG@20 based system rankings for different sets of gains, and two user relevance scenarios: top relevance vs.\ marginal relevance.}
\label{tab:evaluationFedWeb13corr} 
\begin{tabular}{lcccccc}
\hline\noalign{\smallskip}
\textbf{FedWeb13} & top relevance & marginal relevance \\
\noalign{\smallskip}\hline\noalign{\smallskip}
PRM vs.\ binary  & 0.75 & 0.94 \\
PRM vs.\ linear              & 0.96 & 0.92 \\
PRM vs.\ exponential        & 0.99 & 0.93 \\
\noalign{\smallskip}\hline
\end{tabular}
\end{table}

We now consider the TREC FedWeb13 Results Merging Task, in which participants were challenged to design algorithms to create a merged ranking of the top-10 results from 157 online search engines. The official metric was nDCG@20, and for the evaluation, only the first of any returned duplicates was taken into account. We used the same evaluation methods, but altered the gains used for nDCG.
\tabref{tab:evaluationFedWeb13corr} shows Kendall's $\tau$ between rankings of the official results merging runs based on different sets of nDCG gains: binary, PRM, linear, and exponential.  For the top relevance scenario, the difference between the PRM and the binary relevance indicates the necessity of compensating for disagreement ($\tau=0.75$). However, using the PRM, linear, or exponential gains seems to make little difference. In the marginal relevance case, the influence of using the PRM vs.\ binary weights is much smaller ($\tau=0.94$). The PRM-based ranking is still highly correlated with the rankings based on exponential or linear gains, although less than in the top relevance scenario. This is due to the stronger influence of the lower level PRM gains in the marginal relevance scenario.


\section{Conclusions and Future Work} \label{sec:conclusions}
In this paper, we presented and analyzed the Predicted Relevance Model (PRM), which allows evaluating relevance towards a random user instead of purely accepting assessments as ground truth.
The PRM allows quantifying the relevance for a random user, associated with multiple graded or categorical assessment levels, based on the disagreement between assessors.
It was shown how existing evaluation measures can benefit from the PRM, leading to a robust evaluation of search engines with respect to several possible notions of binary user relevance, linked with the assessment levels. In a series of experiments based on existing evaluation collections, we explained how the PRM can be applied in practice, and analyzed its properties in actual evaluation scenarios.

This paper opens up several possibilities for future research. One straightforward direction is in further studying how the PRM can be applied to graded relevance evaluation measures other than the nDCG, or in other scenarios of user relevance. 
Another logical next step is the development of a principled way to combine the original view of graded relevance judgments as a measure of fractional utility, with the PRM ideas based on disagreement probabilities and binary user relevance. 
Furthermore, the PRM covers only one particular aspect of the general pursuit of predicting relevance of results towards users, namely the influence of disagreement. Other aspects that could be taken into account are, for example, the impact of multiple observed judgments per result, characteristics of individual assessors or users, the type of test topics, the result snippet observed by the users, etc.
The relevance of a result given a query, prior to observing one or more assessments, could for example depend on the type of query and the snippet shown to the users. Instead of the disagreement parameters according to the PRM, a more accurate posterior probability of relevance could be calculated after observing the available judgments on that particular result. We hope the insights gained in our current work will help in making progress towards this goal.

\begin{acknowledgements}
First of all, we would like to thank the reviewers. Their particularly detailed comments and suggestions lifted the paper's overall quality and coherence, and played an important role in shaping the formulation and interpretation of the PRM in its current form. We would also like to thank Dolf Trieschnigg for his work on the FedWeb13 data and the many technical discussions, Tetsuya Sakai for providing us with the NTCIR INTENT-2 data, and Bart Deygers for the valuable suggestions to improve the manuscript. 

This work was supported by Ghent University -- iMinds in Belgium, and by the Dutch national program COMMIT and the NWO-Catch project Folktales As Classifiable Texts (FACT) in the Netherlands. 

\end{acknowledgements}

\bibliographystyle{spbasic}      
%
%
%

\end{document}